# Effects of entanglement in an ideal optical amplifier

J.D. Franson and R.A. Brewster
*University of Maryland Baltimore County, Baltimore, MD 21250 USA*

In an ideal linear amplifier, the output signal is linearly related to the input signal with an additive noise that is independent of the input. The decoherence of a quantum-mechanical state as a result of optical amplification is usually assumed to be due to the addition of quantum noise. Here we show that entanglement between the input signal and the amplifying medium can produce an exponentially-large amount of decoherence in an ideal optical amplifier even when the gain is arbitrarily close to unity and the added noise is negligible. These effects occur for macroscopic superposition states, where even a small amount of gain can leave a significant amount of which-path information in the environment. Our results show that the usual input/output relation of a linear amplifier does not provide a complete description of the output state when post-selection is used.

## 1. Introduction

A linear optical amplifier multiplies the input signal by a constant gain $g$ while adding noise that is independent of the input [1-8]. It is generally assumed that all of the degradation of a quantum state that occurs during amplification is due to the addition of quantum noise. Here we show that entanglement between the input signal and the amplifying medium in an ideal optical amplifier will generate "which-path" information that can produce an exponentially-large amount of decoherence even when the gain is arbitrarily close to unity and the added noise is negligibly small. This situation occurs for inputs that are macroscopic superposition states, such as a Schrodinger cat, where even a small amount of gain can result in a significant amount of which-path information left in the environment. Our results show that the usual linear input/output relation of an optical amplifier does not completely describe the output state when post-selection techniques are used to analyze the output.

To be more precise, the output of any linear optical amplifier in the Heisenberg picture is given by [1-8]

$$\hat{x}_{out} = g\hat{x}_{in} + \hat{N}. \qquad (1)$$

Here $\hat{x} \equiv (\hat{a} + \hat{a}^{\dagger})/\sqrt{2}$ is one of the phase quadratures of the signal field, where $\hat{a}$ is the corresponding photon annihilation operator and $\hat{x}_{in}$ and $\hat{x}_{out}$ describe the input and output of the amplifier. $\hat{N}$ is a noise operator that commutes with $\hat{x}_{in}$, and a similar equation describes the other phase quadrature $\hat{p} \equiv (\hat{a} - \hat{a}^{\dagger})/\sqrt{2}i$. The statistical properties of the quantum noise $\hat{N}$ have been analyzed in detail [8] and it is generally considered to be the limiting factor in the performance of an ideal optical amplifier.

In the limit where $g \to 1$, $\hat{N} \to 0$ for an ideal amplifier and Eq. (1) would seem to imply that there should be no significant difference between the input and output fields. That is not the case for macroscopic superposition states as will be shown below. Although Eq. (1) is mathematically correct, $\hat{x}_{out}$ from Eq. (1) cannot be used to calculate the variance and other higher-order moments when post-selection is used. More generally, we will show that the Heisenberg picture approach of Eq (1) is not equivalent to using the Schrodinger picture when non-unitary transformations are applied, as is the case in post-selection.

An optical parametric amplifier (OPA) is a commonly-used example of a linear amplifier, and Caves et al. [8] showed that any ideal (phase-insensitive) linear amplifier can be modeled by an OPA. It is well known that an OPA produces entanglement between the output signal and another optical mode known historically as the idler, as illustrated in Fig. 1. For an OPA, $\hat{N} = -\sqrt{g^2 - 1}\hat{q}_{in}$, where $\hat{q}_{in}$ is the $\hat{x}$ quadrature in the idler mode. The same results apply to any ideal linear amplifier, including those based on an inverted atomic medium [9]. We will use an OPA to illustrate the effects of entanglement on a quantum signal.

The linear relationship of Eq. (1) is only valid in the limit of a strong pump, where the effects of saturation and fluctuations in the pump field are negligible. We will assume throughout that this condition is satisfied and that the pump can be treated classically. A number of earlier papers [10-16] have investigated nonlinear phenomena that can occur when the pump is sufficiently weak that saturation and fluctuations in the pump power are significant, but those effects are unrelated to the decoherence of interest here, which can occur even in an ideal linear amplifier in the limit of a strong pump.

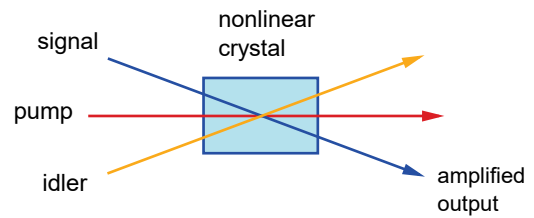

**Fig. 1.** Amplification of a signal by a parametric amplifier. The Hamiltonian corresponds to the annihilation of a photon from the pump beam accompanied by the emission of a photon in both the signal and idler modes. The which-path information produced by the entanglement between the signal and idler modes can produce an exponentially-large amount of decoherence even in the limit of small gain and negligible added noise.

As an example of these effects, we consider the amplification of Schrodinger cat states in the next section. We show that an ideal amplifier can greatly reduce the visibility of the quantum interference between the two terms in a cat state even when the quantum noise is negligible. This situation is analyzed in more detail in Section 3 using the Husimi-Kano Q-function, which allows the visibility of the quantum interference to be calculated analytically. The results from the Q-function calculation in the Schrodinger picture are compared with the results of the Heisenberg picture in Section 4. The implications of these results are discussed in Section 5 along with our conclusions.

## 2. Decoherence of Schrodinger cat states

The decoherence of a Schrodinger cat state by a parametric amplifier will be considered in this section, where the most interesting results correspond to the limit of $g \to 1$. The decoherence of the cat state can be measured using the interferometer arrangement shown in Fig. 2. Earlier studies of the amplification of cat states [4, 17-28] did not consider the limit of $g \to 1$ or the interferometer approach of Fig. 2.

The first step in this process is to generate a Schrodinger cat state by passing a coherent state $|\alpha_0\rangle$ (laser beam) with complex amplitude $\alpha_0$ in the signal mode through a single-photon interferometer that contains a Kerr medium [29] K in one path, as illustrated by the state-preparation box on the left-hand side of Fig. 2. A constant phase shift is applied in such a way that a net phase shift of $\pm\phi$ will be applied to the coherent state depending on the path taken by the single photon $\gamma_1$, as illustrated in Fig. 3b. We post-select on those events in which $\gamma_1$ is detected in the detector labelled $D_1$ in Fig. 2, which produces a Schrodinger cat state [30-32] given by

$$|\psi\rangle = \left(\left|e^{i\phi}\alpha_0\right\rangle + \left|e^{-i\phi}\alpha_0\right\rangle\right)|0_i\rangle/\sqrt{2}. \quad (2)$$

Here we have assumed that the idler mode of the OPA is initially in its vacuum state $|0_i\rangle$. The normalization of Eq. (2) also assumes that $\phi$ is sufficiently large that there is negligible overlap between the two coherent-state components.

The signal mode is then amplified using an OPA with a gain $g = 1+\varepsilon$, which will increase the amplitude of the signal by a relatively small amount for $\varepsilon \ll 1$ as illustrated in Fig. 3c. The amplification will also displace the idler mode in accordance with the relations [7,8]

$$\begin{aligned}\hat{q}_{out} &= g\hat{q}_{in} - \sqrt{g^2-1}\,\hat{x}_{in} \\ \hat{\pi}_{out} &= g\hat{\pi}_{in} + \sqrt{g^2-1}\,\hat{p}_{in},\end{aligned} \quad (3)$$

where $\hat{\pi}$ is the other phase quadrature for the idler. This creates entanglement between the signal and the idler modes, since the idler mode is displaced in different directions in phase space for the two Schrodinger cat state components, as illustrated by the red and blue colors in Fig. 3c. It is important to note that the change in the idler can be much larger than the change in the signal, since $\sqrt{g^2-1} \sim \sqrt{2\varepsilon} \gg \varepsilon$ for $\varepsilon \ll 1$.

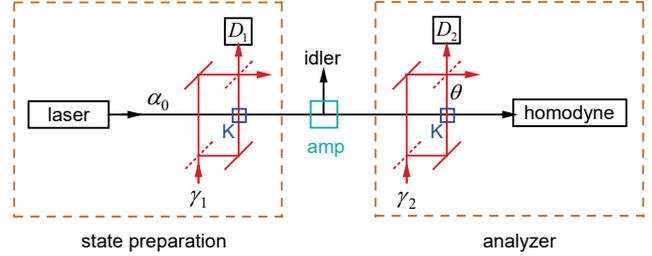

**Fig. 2.** Testing the properties of a parametric amplifier by generating a quantum state, passing it through an amplifier, and then analyzing the properties of the output state. Here a Schrodinger cat state is first produced by passing a coherent state through a single-photon interferometer with a Kerr medium K in one path. After amplification, a second single-photon interferometer will produce quantum interference between the two components of the cat state when a homodyne measurement indicates a net phase shift near zero, as illustrated in Fig. 3. This allows a measurement of the amount of decoherence due to entanglement between the signal and idler modes in the amplifier, which can occur even when the quantum noise is negligibly small. Here $\gamma_1$ and $\gamma_2$ represent single photons, $D_1$ and $D_2$ are single-photon detectors, and the pump beam for the parametric amplifier is not shown.

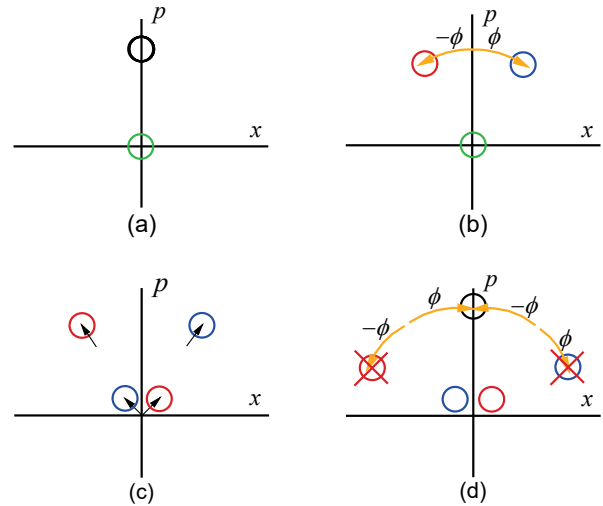

**Fig. 3.** Qualitative phase-space description of the interferometer system of Fig. 2, where x and p correspond to the two quadratures of the fields. (a) The initial state in which the signal mode is a coherent state with amplitude $\alpha_0$ as represented by the black circle, while the idler mode is in its vacuum state represented by the green circle. (b) The state of the system after the first single-photon interferometer, where there are equal probability amplitudes that the signal mode has been shifted in phase by $\pm\phi$. (c) Entangled state created by the parametric amplifier of Fig. 1, where the signal and idler modes have been displaced in correlated directions. (d) The state of the system after the final single-photon interferometer, where a second phase shift of $\pm\phi$ can recombine the two components of the Schrodinger cat. The visibility of the quantum interference between these two probability amplitudes is reduced exponentially by the remaining entanglement with the idler mode.

After the amplifier, the coherence properties of the output signal are analyzed using another single-photon interferometer with a Kerr medium as shown in the analyzer box on the right-hand side of Fig. 2. This process also applies a phase shift of $\pm\phi$, where we post-select on single photon $\gamma_2$ having been detected in $D_2$. The net phase shift after the second interferometer will be either 0 or $\pm 2\phi$ as indicated by the arrows in Fig. 3d. The phase of the final signal mode is measured using a homodyne detector and we post-select on those events in which there was a net phase shift near 0, as indicated by the red crosses in Fig. 3d. This can only occur if there was one positive and one negative phase shift.

If we ignore the effects of any amplification for the time being (g=1), this process results in a post-selected state $|\psi'\rangle$ given by

$$|\psi'\rangle = \left(|\alpha'_+\rangle + e^{i\theta}|\alpha'_-\rangle\right)|0_i\rangle / 2^{3/2}. \tag{4}$$

The state $|\alpha'_+\rangle$ corresponds to a coherent state $|\alpha_0\rangle$ whose phase was initially shifted by $+\phi$ by the first single-photon interferometer and then shifted by $-\phi$ in the second single-photon interferometer, as illustrated in Fig. 2. The two phase shifts cancel out to give back the initial coherent state, so that $|\alpha'_+\rangle = |\alpha_0\rangle$. In a similar way, $|\alpha'_-\rangle$ corresponds to a coherent state $|\alpha_0\rangle$ whose phase was initially shifted by $-\phi$ by the first single-photon interferometer and then shifted by $+\phi$ in the second single-photon interferometer, so that $|\alpha'_-\rangle = |\alpha_0\rangle$ as well. The parameter $\theta$ in Eq. (4) is a phase shift applied to photon $\gamma_2$ in one path of the second interferometer, as illustrated in Fig. 2.

Eq. (4) corresponds to the amplitudes of the post-selected terms in the state vector without any renormalization. The probability $P$ of such a sequence of events occurring is thus equal to the norm of $|\psi'\rangle$, which reduces to

$$P = \frac{1}{2}\cos^2(\theta/2). \tag{5}$$

We can define the visibility $v$ of the quantum interference as usual by $v \equiv (P_{max} - P_{min})/(P_{max} + P_{min})$, where $P_{max}$ and $P_{min}$ are the maximum and minimum probabilities obtained by varying the phase $\theta$. Thus the probability $P$ depends sinusoidally on $\theta$ and there is 100% visibility of the quantum interference in the absence of any amplification. The entire system can be viewed as a Schrodinger-cat interferometer.

We now consider the effect of amplification in the optical path between the two single-photon interferometers as shown in Fig. 2. To a first approximation, the entanglement between the signal and idler modes has the effect of replacing Eq. (4) with

$$|\psi''\rangle = \left(|\alpha''_+\rangle|i_+\rangle + e^{i\theta}|\alpha''_-\rangle|i_-\rangle\right)/2^{3/2}, \tag{6}$$

where $|\alpha''_\pm\rangle$ is the amplified signal and $|i_\pm\rangle$ is the corresponding state of the idler mode as illustrated in Fig. 3d. Now the cross-terms in $\langle\psi''|\psi''\rangle$ that are responsible for the quantum interference will be proportional to $\langle i_+|i_-\rangle$. This will produce an exponential decrease in the visibility when the gain is sufficiently high that there is very little overlap between the two displaced states of the idler mode.

The most interesting situation occurs when $\varepsilon << 1$ but $|\alpha_0|$ is so large that $\varepsilon|\alpha_0|^2 >> 1$. In that limit, the added noise $\hat{N} = -\sqrt{g^2-1}\hat{q}_{in} \to 0$ and $\hat{x}_{out} = \hat{x}_{in}$ from Eq. (1). It can be shown that $\hat{p}_{out} = \hat{p}_{in}$ as well. As a result, one might expect that there should be no significant change in the field due to the amplifier. But the magnitude of the displacement of the idler mode from Eq. (3) is given by $\sqrt{g^2-1}|\alpha_0| \sim \sqrt{2\varepsilon}|\alpha_0| >> 1$. Because the displacement of the idler in phase space is much larger than the width of its gaussian distribution, the idler overlap $\langle i_+|i_-\rangle$ will be exponentially small and the visibility will approach zero in that limit, in contrast to what might be expected from Eq. (1).

Although $|\alpha_0|$ was assumed to be relatively large in the example above, the amplitude of the pump field can be even larger so that saturation and fluctuations in the pump are negligible and the linear input/output relation of Eqs (1) and (3) are satisfied. This will be the case provided that the number of photons in the pump field is much larger than the number of signal and idler photons that are emitted. As we will show in more detail in the next section, the decoherence of interest will be exponentially small even if only a few idler photons are emitted and large pump intensities are not actually required.

### 3. Analysis using the Q-function

Eq. (6) is only approximate because it assumes a product state between the corresponding signal and idler modes. The effects of the amplification can be calculated exactly using the two-mode Husimi-Kano quasiprobability distribution [33-35] defined by

$$Q(\alpha,\beta) \equiv \frac{1}{\pi^2}\langle\alpha|\langle\beta|\hat{\rho}|\beta\rangle|\alpha\rangle. \tag{7}$$

Here $\hat{\rho}$ is the density operator for the system, $|\alpha\rangle$ and $|\beta\rangle$ denote arbitrary coherent states in the signal and idler modes, and $|\alpha\rangle|\beta\rangle$ is short-hand notation for $|\alpha\rangle\otimes|\beta\rangle$.

After the first single-photon interferometer and before the amplifier, the density matrix corresponding to the pure state of Eq. (2) is given by

$$\hat{\rho} = \hat{\rho}_{++} + \hat{\rho}_{+-} + \hat{\rho}_{-+} + \hat{\rho}_{--}, \tag{8}$$

where

$$\hat{\rho}_{++} = |\psi_+\rangle\langle\psi_+| = |e^{i\phi}\alpha_0\rangle|0_i\rangle\langle 0_i|\langle e^{i\phi}\alpha_0|/2$$
$$\hat{\rho}_{+-} = |\psi_+\rangle\langle\psi_-| = |e^{i\phi}\alpha_0\rangle|0_i\rangle\langle 0_i|\langle e^{-i\phi}\alpha_0|/2$$
$$\hat{\rho}_{-+} = |\psi_-\rangle\langle\psi_+| = |e^{-i\phi}\alpha_0\rangle|0_i\rangle\langle 0_i|\langle e^{i\phi}\alpha_0|/2 \quad (9)$$
$$\hat{\rho}_{--} = |\psi_-\rangle\langle\psi_-| = |e^{-i\phi}\alpha_0\rangle|0_i\rangle\langle 0_i|\langle e^{-i\phi}\alpha_0|/2.$$

Here $|\psi_\pm\rangle$ corresponds to the two states in Eq. (2), where the $\pm$ signs will always refer to the sign of the phase shift applied in the first interferometer.

The Q-function defined by Eq. (7) can therefore be written as

$$Q(\alpha,\beta) = Q_{++}(\alpha,\beta) + Q_{+-}(\alpha,\beta) + Q_{-+}(\alpha,\beta) + Q_{--}(\alpha,\beta), \quad (10)$$

where

$$Q_{+-}(\alpha,\beta) \equiv \frac{1}{\pi^2}\langle\alpha|\langle\beta|\hat{\rho}_{+-}|\beta\rangle|\alpha\rangle, \quad (11)$$

for example. The cross terms $Q_{+-}(\alpha,\beta)$ and $Q_{-+}(\alpha,\beta)$ are responsible for the quantum interference and will be of particular interest.

The effects of the amplifier on the Q-function can be calculated from the interaction Hamiltonian $\hat{H}'$ of an OPA, which is given by [7,8]

$$\hat{H}' = i\hbar\kappa(\hat{a}\hat{b} - \hat{a}^\dagger\hat{b}^\dagger). \quad (12)$$

Here $\hat{a}$ and $\hat{b}$ are the annihilation operators for the signal and idler modes, respectively, while $\kappa$ is a parameter that reflects the strength of the interaction. The time evolution operator $\hat{S}(r)$ for the system can be shown to have the form [8]

$$\hat{S}(r) \equiv e^{-i\hat{H}'t/\hbar} = e^{r(\hat{a}\hat{b} - \hat{a}^\dagger\hat{b}^\dagger)}. \quad (13)$$

Here $t$ is the interaction time and the parameter $r$ is defined by $r = \kappa t$. Eq. (13) can be put into the factored form [36]

$$\hat{S}(r) = \frac{1}{g}e^{-\sqrt{g^2-1}\hat{a}^\dagger\hat{b}^\dagger/g}g^{-(\hat{a}^\dagger\hat{a}+\hat{b}^\dagger\hat{b})}e^{\sqrt{g^2-1}\hat{a}\hat{b}/g}, \quad (14)$$

where the gain $g$ is defined as $g = \cosh(r)$.

The density matrix $\hat{\rho}'$ after the amplification process is given by $\hat{\rho}' = \hat{S}(r)\hat{\rho}\hat{S}^\dagger(r)$. Combining this with Eqs. (7) through (9) gives

$$Q_{+-}(\alpha,\beta) = \frac{1}{2\pi^2}\big(\langle\alpha|\langle\beta|\hat{S}(r)|0_i\rangle|\alpha_+\rangle\big) \times \big(\langle\alpha_-|\langle 0_i|\hat{S}^\dagger(r)|\beta\rangle|\alpha\rangle\big). \quad (15)$$

Here we have defined $|\alpha_+\rangle = |e^{i\phi}\alpha_0\rangle$ and $|\alpha_-\rangle = |e^{-i\phi}\alpha_0\rangle$. Using Eq. (14) and the fact that $\hat{a}|\alpha\rangle = \alpha|\alpha\rangle$, $\hat{b}|\beta\rangle = \beta|\beta\rangle$, and $\langle\beta|0\rangle = \exp[-|\beta|^2/2]$, the first factor $f_+$ on the right-hand side of Eq. (15) can be rewritten as

$$f_+ \equiv \big(\langle\alpha|\langle\beta|\hat{S}(r)|0_i\rangle|\alpha_+\rangle\big) = \frac{1}{g}e^{-|\beta|^2/2}e^{-\sqrt{g^2-1}\alpha^*\beta^*/g} \times \langle\alpha|\big(g^{-\hat{a}^\dagger\hat{a}}\big)|\alpha_+\rangle. \quad (16)$$

By expanding the coherent states $|\alpha\rangle$ and $|\alpha_+\rangle$ in the number-state basis, it can be shown that Eq. (16) is equivalent to

$$f_+ = \frac{1}{g}e^{-|\alpha|^2/2}e^{-|\alpha_+|^2/2}e^{-|\beta|^2/2}e^{-\sqrt{g^2-1}\alpha^*\beta^*/g}e^{\alpha^*\alpha_+/g}. \quad (17)$$

A similar result can be obtained for the other factor $f_-$ on the right-hand side of Eq. (15). Inserting these results back into Eq. (15) gives the Q-function after the amplification process in the form

$$Q_{+-}(\alpha,\beta) = \frac{1}{2\pi^2 g^2}\left(e^{-|\beta|^2}e^{-\sqrt{g^2-1}(\alpha^*\beta^* + \alpha\beta)/g}\right) \times \left(e^{-|\alpha|^2}e^{-|\alpha_+|^2/2}e^{-|\alpha_-|^2/2}e^{(\alpha^*\alpha_+ + \alpha\alpha_-^*)/g}\right). \quad (18)$$

The same procedure can be used to find the three other terms in the Q-function of Eq. (10).

The Kerr medium in the second two-photon interferometer after the amplifier will shift the phase of the signal mode by $\pm\phi$, which is equivalent to applying a transformation $\hat{T}$ given by

$$\hat{T} = \frac{1}{2}\big(e^{i\theta}\hat{U}_+ + \hat{U}_-\big), \quad (19)$$

where $\hat{U}_\pm$ applies a phase shift of $\pm\phi$ to the signal. The factor $f_+$ in Eq. (16) will then be transformed into

$$f_+ = \frac{1}{g}e^{-|\beta|^2/2}\langle\alpha|\hat{T}\left(e^{-\sqrt{g^2-1}\hat{a}^\dagger\beta^*/g}g^{-\hat{a}^\dagger\hat{a}}\right)|\alpha_+\rangle \quad (20)$$

Letting the operator $\hat{T}^\dagger$ act to the left on the state $\langle\alpha|$ just shifts its phase accordingly. Using a similar procedure on the other factor $f_-$ and dropping the two phase-shifted terms that are rejected in the post-selection process gives the final Q-function:

$$Q_{+-}(\alpha,\beta) = \frac{1}{4\pi^2 g^2}\left\{e^{-|\beta|^2}\exp\left[-\sqrt{g^2-1}(e^{-i\phi}\alpha^*\beta^* + e^{-i\phi}\alpha\beta)/g\right]\right\}$$
$$\times \left(e^{-|\alpha|^2}e^{-|\alpha_0|^2}e^{(\alpha^*\alpha_0+\alpha\alpha_0^*)/g}\right)e^{-i\theta},$$
(21)

with similar expressions for the three other terms.

The probability $P$ of a post-selected event of this kind is given by

$$P = \int d^2\alpha \, d^2\beta \, Q(\alpha,\beta). \quad (22)$$

An examination of Eq. (21) shows that the last factor in parentheses involving $\alpha_0$ is strongly peaked when the value of $\alpha$ is near $g\alpha_0$, as would be expected from Fig. 3d. The value of the first factor involving $\beta$ is exponentially small for those values of $\alpha$ due to the phase shift of $\pm\phi$ that remains in the idler mode. As a result, the $Q_{+-}$ and $Q_{-+}$ terms are greatly reduced and the visibility of the interference becomes exponentially small for sufficiently large gain.

Combining Eqs. (21) and (22) and performing the integrals gives the probability $P_{+-}$ associated with $Q_{+-}$, with similar results for the other components. This can be used to evaluate the visibility $v \equiv (P_{max} - P_{min})/(P_{max} + P_{min})$, where $P_{max}$ and $P_{min}$ are the maximum and minimum probabilities obtained by varying the phase $\theta$. The result is

$$v = \frac{e^{-2|\alpha_0|^2(g^2-1)/(2g^2-1)}}{(2g^2-1)} \to \exp(-4\varepsilon|\alpha_0|^2), \quad (23)$$

where we have taken $\phi = \pi/2$ and the limit on the right-hand side corresponds to $\varepsilon \ll 1$. If we choose $\varepsilon$ and $|\alpha_0|$ such that $\varepsilon|\alpha_0| \ll 1$ but $\varepsilon|\alpha_0|^2 \gg 1$, then $(g-1)|\langle\hat{x}_{in}\rangle| \sim \varepsilon|\alpha_0| \ll 1$ and $\hat{x}_{out} - \hat{x}_{in} \to 0$ from Eq. (1). ($\hat{p}_{out} = \hat{p}_{in}$ as well). Thus the entanglement between the signal and the idler can give an exponentially-small visibility even when the difference between the input and output signals would be arbitrarily small according to Eq. (1). The visibility is 100% in the opposite limit where $|\alpha_0|$ is held constant as $g \to 1$.

We considered the limit of $\varepsilon|\alpha_0|^2 \gg 1$ in the discussion above in order to show that the added noise is not responsible for these effects. A large value of $|\alpha_0|$ in a Schrodinger cat state is currently not feasible experimentally. Eq. (23) can still be tested under more relaxed conditions, such as $\varepsilon = 1/2$ and $|\alpha_0| = 1$ for example, which would also produce a large amount of decoherence due to entanglement with the idler. Experiments of this kind appear to be difficult but feasible using current technology.

It is well known that macroscopic superposition states are very susceptible to decoherence and one might argue that it is not surprising that a large amount of decoherence occurs in the limit of $\varepsilon|\alpha_0|^2 \gg 1$. What is surprising is that Eq. (1) does not provide an adequate description of this decoherence for reasons that will be discussed in section 4. In fact, Eq. (1) seems to suggest that there should be no significant change in the signal in this limit where $\hat{x}_{out} = \hat{x}_{in}$, while Eq. (23) predicts a large amount of decoherence even in that case.

Although Eq. (1) is correct, the operator $\hat{x}$ (or $\hat{p}$) does not represent the measurement outlined in Figs. 2 and 3. A single mode of the electromagnetic field is mathematically equivalent to a simple harmonic oscillator, and the post-selected measurements of interest here have an expectation value given in the coordinate representation [37] that is roughly analogous to

$$\langle \hat{O} \rangle = \int |\psi(x_1) + e^{i\theta}\psi(x_2)|^2 \, dx. \quad (24)$$

Here $\hat{O}$ is the corresponding operator and $\psi(x)$ is the wave function, while $x_1$ and $x_2$ are two different locations. The cross-terms in Eq. (24) produce interference effects that are dependent on the relative phase $\theta$. Eq. (24) is very different from the expectation value of the operator $\hat{x}$, which is independent of the relative phase of the wave function at two different points. This example reflects the fact that a knowledge of $\hat{x}$ does not determine all of the observable properties of the system. As will be discussed in the following section, the variance and higher moments of the distribution cannot be derived from operator $\hat{x}$ in the Heisenberg picture as a result of the post-selection process.

In a similar way, Caves et al. [8] traced over the idler to obtain the Q-function $\tilde{Q}(\alpha)$ for the signal mode alone:

$$\tilde{Q}(\alpha) \equiv \int d^2\beta \, Q(\alpha,\beta). \quad (25)$$

They showed that the effects of an OPA on $\tilde{Q}(\alpha)$ are given by

$$\tilde{Q}_{out}(\alpha) = \tilde{Q}_{in}(\alpha/g)/g^2. \quad (26)$$

Eq. (26) has a remarkably simple form and one might infer from it that there is no significant change in the signal in the limit of $g \to 1$, since $\tilde{Q}_{out}(\alpha) - \tilde{Q}_{in}(\alpha) \to 0$ in the limit discussed above. Nevertheless, Eq. (26) is correct and it can be used to derive the same visibility as that of Eq. (23). This shows that very small changes in the Q-function can have surprisingly large effects on the outcome of an experiment.

The importance of the idler mode in a parametric amplifier is well known, but it is usually assumed that its effects are completely described by the additive noise $\hat{N}$. For example, Caves et al. [8] recently analyzed linear amplifier noise in detail, with an emphasis on calculating the higher-order moments of the added noise distribution. As they put it, "the amplification of the primary mode requires it to be coupled to other physical systems, not least to provide the energy needed for amplification; these other systems, which can be thought of as the amplifier's internal degrees of freedom, necessarily add noise to the output." We certainly agree with that statement, but our point is that the effects of entanglement with the idler are not limited to the production of an additive noise $\hat{N}$. The amplification of Schrodinger cats was considered long ago by R. Glauber [4], but he also concluded that the decoherence was due to the addition of noise.

## 4. Comparison with the Heisenberg picture

The Q-function calculations in the previous section were performed in the Schrodinger picture, while the usual input-output relations of Eq. (1) are based on the Heisenberg picture. In this section, we will show that these two approaches are not equivalent when post-selection is used.

The limitations in the use of the Heisenberg picture to describe the Schrodinger cat interferometer of Fig. 2 can be understood by calculating the variance of $\hat{x}_F$, the quadrature of the final output field that enters the homodyne detector. Consider the case in which we post-select on the output of the two single-photon detectors shown in Fig. 2, but with no post-selection based on the output of the homodyne detector. The probability of measuring a coherent-state phase shift of $\pm 2\phi$ is independent of $\theta$, while the probability of measuring a total phase shift of 0 depends on $\cos^2(\theta/2)$ in the absence of decoherence as in Eq. (5). As a result, the variance $\langle \hat{x}_F^2 \rangle$ will also depend on $\theta$ and it can be used to compare the two approaches; for simplicity, we will consider the case where $\langle \hat{x}_F \rangle = 0$.

In the Schrodinger picture, the final state $|\psi_F\rangle$ of the system is given by

$$|\psi_F\rangle = \hat{T}\hat{S}|\psi_o\rangle. \quad (27)$$

Here the initial state $|\psi_0\rangle$ corresponds to a Schrodinger cat state, while $\hat{S} = \hat{S}(r)$ is the transformation produced by the OPA and $\hat{T}$ is the transformation produced by the second single-photon interferometer and the post-selection process. For simplicity, we define a new operator $\hat{O} = \hat{T}\hat{S}$, which allows Eq. (27) to be rewritten as

$$|\psi_F\rangle = \hat{O}|\psi_o\rangle. \quad (28)$$

It should be noted that the operators $\hat{T}$ and $\hat{O}$ are not unitary, which requires some care in the use of the Heisenberg picture.

The expectation value of the Schrodinger operator $\hat{x}_S$ at the output of the interferometer is then given by

$$\langle \psi_F | \hat{x}_S | \psi_F \rangle = \langle \hat{O}\psi_0 | \hat{x}_S | \hat{O}\psi_0 \rangle = \langle \psi_0 | \hat{O}^\dagger \hat{x}_S \hat{O} | \psi_0 \rangle. \quad (29)$$

If we define the Heisenberg operator $\hat{x}_F$ as usual by $\hat{x}_F = \hat{O}^\dagger \hat{x}_S \hat{O}$, then

$$\langle \psi_F | \hat{x}_S | \psi_F \rangle = \langle \psi_0 | \hat{x}_F | \psi_0 \rangle. \quad (30)$$

Thus the usual definition of the Heisenberg operator $\hat{x}_F$ gives the correct expectation value even though the operators $\hat{T}$ and $\hat{O}$ are not unitary.

The situation is not so simple for the variance, however. In the Schrodinger picture, this can be written as

$$\langle \psi_F | \hat{x}_S^2 | \psi_F \rangle = \langle \psi_0 | \hat{O}^\dagger \hat{x}_S^2 \hat{O} | \psi_0 \rangle = \langle \psi_0 | \hat{O}^\dagger \hat{x}_S \hat{O} \hat{O}^{-1} \hat{x}_S \hat{O} | \psi_0 \rangle, \quad (31)$$

where we have used the fact that $\hat{O}\hat{O}^{-1} = \hat{I}$ and we have assumed that $\langle \psi_F | \hat{x}_S | \psi_F \rangle = 0$ for simplicity. By comparison, the expectation value of the operator $\hat{x}_F^2$ in the Heisenberg picture is given by

$$\langle \psi_0 | \hat{x}_F^2 | \psi_0 \rangle = \langle \psi_0 | \hat{O}^\dagger \hat{x}_S \hat{O} \hat{O}^\dagger \hat{x}_S \hat{O} | \psi_0 \rangle. \quad (32)$$

A comparison of Eqs. (31) and (32) shows that

$$\langle \psi_0 | \hat{x}_F^2 | \psi_0 \rangle \neq \langle \psi_F | \hat{x}_S^2 | \psi_F \rangle, \quad (33)$$

since $\hat{O}^\dagger \neq \hat{O}^{-1}$ if operator $\hat{O}$ is not unitary. Thus the Heisenberg operator $\hat{x}_F^2$ does not give the correct variance for the Schrodinger cat interferometer of interest here, even though $\hat{x}_F$ does give the correct expectation value in Eq. (30).

To show this more explicitly, we will calculate the variance predicted by the use of Eq. (1) in the Heisenberg picture. From Eq. (27), $\hat{x}_F$ can be also be written as

$$\hat{x}_F = \hat{S}^\dagger \hat{T}^\dagger \hat{x}_S \hat{T}\hat{S}. \quad (34)$$

If we ignore the fact that $\hat{T}$ is not unitary for the time being, we can use Eq. (1) to evaluate the results of the linear transformation of the operator $\hat{T}^\dagger \hat{x}_S \hat{T}$ in Eq. (34):

$$\hat{x}_F = \hat{S}^\dagger \left( \hat{T}^\dagger \hat{x}_S \hat{T} \right) \hat{S} = g\left( \hat{T}^\dagger \hat{x}_S \hat{T} \right) - \sqrt{g^2-1}\left( \hat{T}^\dagger \hat{q}_S \hat{T} \right). \quad (35)$$

Eq. (35) is equivalent to having amplified an input signal with a quadrature given by $\hat{T}^\dagger \hat{x}_S \hat{T}$.

Eq. (35) can now be used to calculate the variance in the Heisenberg picture, which gives

$$\langle \psi_0 | \hat{x}_F^2 | \psi_0 \rangle = \langle \psi_0 | \left[ g\left( \hat{T}^\dagger \hat{x}_S \hat{T} \right) - \sqrt{g^2-1}\left( \hat{T}^\dagger \hat{q}_S \hat{T} \right) \right]^2 | \psi_0 \rangle. \quad (36)$$

Expanding the square in Eq. (36) gives

$$\langle \hat{x}_F{}^2 \rangle = g^2 \langle \psi_0 | (\hat{T}^\dagger \hat{x}_S \hat{T})^2 | \psi_0 \rangle$$
$$+ (g^2 - 1)\langle \psi_0 | (\hat{T}^\dagger \hat{q}_S \hat{T})^2 | \psi_0 \rangle, \quad (37)$$

where we have made use of the fact that $\hat{q}_S$ is uncorrelated with the other term in Eq. (36). The operator $\hat{x}_W = \hat{T}^\dagger \hat{x}_S \hat{T}$ can be identified as the input to the homodyne detector in the absence of any amplification. Thus the variance in Eq. (37) is on the order of

$$\langle \hat{x}_F{}^2 \rangle \approx g^2 \langle \hat{x}_W{}^2 \rangle + (g^2 - 1). \quad (38)$$

Eq. (38) makes use of the fact that $\langle \hat{q}_S{}^2 \rangle = 1$ and $\hat{T}$ is just a phase shift. The same analysis that led to Eq. (5) gives a post-selected variance of $\langle \hat{x}_W{}^2 \rangle = \sin^2(2\phi) |\alpha_0|^2 / (1 + 2\cos^2\theta)$ without amplification, where we have assumed $\text{Re}(\alpha_0) = 0$ and $|\alpha_0| \gg 1$ for simplicity.

The Heisenberg picture results of Eq. (38) predict that $\langle \hat{x}_F{}^2 \rangle$ will be unaffected by the OPA provided that $g \ll 1$, regardless of the value of $|\alpha_0|^2$. Thus there would be a strong dependence on the value of $\theta$ even when $\varepsilon |\alpha_0|^2 \gg 1$ provided that $\varepsilon$ itself is small, which is in disagreement with Eq. (23) from the Schrodinger picture.

This result can be understood intuitively from the fact that the addition of a small amount of noise to the output signal would have the same effect as a small amount of phase noise added to the local oscillator used in the homodyne detector; both simply shift the measured value of $\hat{x}$ by a small amount. That does not include the effects of entanglement, and Eq. (23) from the Q-function analysis in the Schrodinger picture predicts an exponentially-small dependence on $\theta$ under these conditions.

The use of the linear input/output relation of Eq. (1) is not valid in the above calculation because the operator $\hat{T}$ is not unitary. A unitary transformation preserves all of the usual commutation relations such as $[\hat{p}, \hat{x}] = \hbar/i$, which are an essential part of the derivation of Eq. (1). The commutation relations are not preserved by a non-unitary transformation and the use of Eq. (1) is no longer valid. Thus the usual linear input/output relations cannot be used to calculate the correct variance if post-selection is used.

The Heisenberg picture has the advantage that it results in a set of differential equations for the quadrature operators that can be solved analytically to give the usual linear input/output relations for $\hat{x}$ and $\hat{q}$. Unfortunately, those results for the quadratures themselves cannot be used to correctly calculate higher moments such as the variance in the example considered here.

## 5. Discussion and conclusions

We have shown that entanglement between the signal and the amplifying medium can produce an exponentially-large amount of decoherence in the amplification of macroscopic superposition states even when the added noise is negligibly small. This shows that the added noise $\hat{N}$ is not totally responsible for the decoherence of an arbitrary amplifier input state, as is commonly assumed.

The linear input/output relation of Eq. (1) seems to imply that there should be no change in the quantum state in the limit of $g \to 1$ and $\hat{N} \to 0$, and this is confirmed by the predicted variance in Eq. (38). That is not the case as we have shown, and the effects of entanglement with the idler are not limited to the addition of noise. Although Eq. (1) is mathematically correct, it does not provide a complete description of the properties of the output state since the variance and other higher moments are not related in any simple way to the Heisenberg operators $\hat{x}_F$ and $\hat{q}_F$ when post-selection is applied. Higher-order moments are required in order to describe correlations and other effects associated with entanglement, and this capability is what is missing in the usual linear input-output relation of Eq. (1).

It is well known that Schrodinger cat states are very susceptible to decoherence, and similar results could be obtained for other macroscopic superposition states. Decoherence will also occur if a cat state is passed through a beam splitter, for example [37]. In that case, which-path information is left in the other output port of the beam splitter in analogy with the information left in the idler mode in this example. Thus decoherence due to the generation of which-path information is not unique to amplifiers, but it cannot be described by the addition of noise in either case. For example, the decoherence of a macroscopic superposition state by a beam splitter cannot be understood on the basis of vacuum fluctuations coupled in from the other input port.

Decoherence is one of the most important issues in quantum information processing and a fundamental understanding of its origins is essential. We previously described a generalization of the interferometer in Fig. 2 in which two coherent states become entangled in phase and violate Bell's inequality [37]. Systems of that kind may be useful for quantum communications and quantum computing, and optical amplifiers may play an important role in the presence of detector noise. Optical amplifiers can enhance the performance of quantum sensor systems as well. Thus the effects of entanglement in optical amplifiers may be of practical importance as well as providing additional insight into the nature of amplifiers.

## Acknowledgements

This work was supported in part by the National Science Foundation under grant No. 1402708.

## References

1. A. Haus and J.A. Mullen, "Quantum noise in linear amplifiers", Phys. Rev. **128**, 2407 (1962).
2. B.R. Mollow and J. Glauber, "Quantum theory of parametric amplification. I", Phys. Rev. **160**, 1076 (1967).
3. C.M. Caves, "Quantum limits on noise in linear amplifiers", Phys. Rev. D **26**, 1817 (1982).
4. R.J. Glauber, "Amplifiers, Attenuators, and Schrodinger's Cat", in *New Techniques and Ideas in Quantum*

*Measurement Theory*, D.M. Greenberger, ed., Annals of the New York Academy of Science **480**, 336 (1986).
5. S. Stenholm, "The theory of quantum amplifiers", Physica Scripta **T12**, 56 (1986).
6. G.S. Agarwal and K. Tara, "Transformations of the nonclassical states by an optical amplifier", Phys. Rev. A **47**, 3160 (1993).
7. A.A. Clerk, M.H. Devoret, S.M. Girvin, F. Marquardt, and R.J. Schoelkopf, "Introduction to quantum noise, measurement, and amplification", Rev. Mod. Phys. **82**, 1155 (2010).
8. C.M. Caves, J. Combes, Z. Jiang, and S. Pandey, "Quantum limits on phase-preserving linear amplifiers", Phys. Rev. A **86**, 063802 (2012).
9. J.D. Franson and B.T. Kirby, "Origin of Quantum Noise and Decoherence in Distributed Amplifiers", Phys. Rev. A **92**, 053825 (2015).
10. R. Graham and H. Haken, "The quantum-fluctuations of the optical parametric oscillator. I", Z. Physik **210**, 276 (1968).
11. R. Graham, "Photon statistics of the optical parametric oscillator including the threshold region", Z. Physik, **210**, 319 (1968).
12. S. Chaturvedi, K. Dechoum, and P.D. Drummond, "Limits to squeezing in the degenerate optical parametric oscillator", Phys. Rev. A **65**, 033805 (2002).
13. P.D. Drummond, K. Dechoum, and S. Chaturvedi, "Critical quantum fluctuations in the degenerate parametric oscillator", Phys. Rev. A **65**, 033806 (2002).
14. K. Dechoum, P.D. Drummond, S. Chaturvedi, and M.D. Reid, "Critical fluctuations and entanglement in the nondegenerate parametric oscillator", Phys. Rev. A **70**, 053807 (2004).
15. S. Chaturvedi and P.D. Drummond, "Stochastic diagrams for critical point spectra", Eur. Phys. J. B **8**, 251 (1999).
16. V. D'Auria, C. de Lisio, A. Porzio, S. Solimeno, J. Anwar, and M.G.A. Paris, "Non-Gaussian states produced by close-to-threshold optical parametric oscillators: Role of classical and quantum fluctuations", Phys. Rev. A **81**, 033846 (2010).
17. D.J. Daniel and G.J. Milburn, "Destruction of quantum coherence in a nonlinear oscillator via attenuation and amplification", Phys. Rev. A **39**, 4628 (1989).
18. V. Buzek, M.S. Kim, and Ts. Gantsog, "Quantum phase distributions of amplified Schrodinger-cat states of light", Phys. Rev. A **48**, 3394 (1993).
19. M.S. Kim, K.S. Lee, and V. Buzek, "Amplification of superposition states in phase-sensitive amplifiers", Phys. Rev. A **47**, 4302 (1993).
20. U. Leonhardt, "Quantum statistics of a two-mode SU(1,1) interferometer", Phys. Rev. A **49**, 1231 (1994).
21. G.M. D'Ariano, M. Fortunato, and P. Tombesi, "Isotropic phase number squeezing and macroscopic quantum coherence", Il Nuovo Cim. B **110**, 1127 (1995).
22. H. Huang, S.-Y. Zhu, and M.S. Zubairy, "Preservation of nonclassical character during the amplification of a Schrodinger cat state", Phys. Rev. A **53**, 1027 (1996).
23. G.S. Agarwal, "Mesoscopic superpositions of states: Approach to classicality and diagonalization in a coherent state basis", Phys. Rev. A **59**, 3071 (1999).
24. M.S. Zubairy and S. Qamar, "Observing the quantum interference using phase-sensitive amplification", Optics Comm. **179**, 275 (2000).
25. R. Filip and J. Perina, "Amplification of Schrodinger-cat state: Distinguishability and interference in phase space", J. Opt. B: Quantum Semiclass. Opt. **3**, 21 (2001).
26. R. Filip, "Amplification of Schrodinger-cat state in a degenerate optical parametric amplifier", J. Opt. B: Quantum Semiclass. Opt. **3**, p. S1 (2001).
27. V.V. Dodonov, C. Valverde, L.S. Souza, and B. Baseia, "Classicalization times of parametrically amplified "Schrodinger cat" states coupled to phase-sensitive reservoirs", Phys. Lett. A **375**, 3668 (2011).
28. A. Laghaout, J.S. Neergaard-Nielsen, I. Rigas, C. Kragh, A. Tipsmark, and U.L. Andersen, "Amplification of realistic Schrodinger-cat-like states by homodyne heralding", Phys. Rev. A **87**, 043826 (2013).
29. R.B. Boyd, *Nonlinear Optics* (Academic Press, San Diego, 2003).
30. B.C. Sanders, "Entangled coherent states", Phys. Rev. A **45**, 6811 (1992).
31. C.C. Gerry, "Generation of optical macroscopic quantum superposition states via state reduction with a Mach-Zehnder interferometer containing a Kerr medium", Phys. Rev. A **59**, 4095 (1999); C.C. Gerry and R. Grobe, "Nonlocal entanglement of coherent states, complementarity, and quantum erasure", Phys. Rev. A **75**, 034303 (2007).
32. W. Schleich, M. Pernigo, and F.L. Kien, "Nonclassical state from two pseudoclassical states", Phys. Rev. A **44**, 2172 (1991).
33. K. Husimi, "Some formal properties of the density matrix", Proc. Phys. Math. Soc. Jpn. **22**, 264 (1940).
34. Y. Kano, "A new phase-space distribution function in the statistical theory of the electromagnetic field", J. Math. Phys. **6**, 1913 (1965).
35. W.P. Schleich, *Quantum Optics in Phase Space* (WILEY-VCH, Berlin, 2001).
36. B.L. Schumaker and C.M. Caves, "New formalism for two-photon quantum optics. II. Mathematical foundation and compact notation", Phys. Rev. A **31**, 3093 (1985).
37. B.T. Kirby and J.D. Franson, "Nonlocal interferometry using macroscopic coherent states and weak nonlinearities", Phys. Rev. A **87**, 053822 (2013); "Macroscopic state interferometry over large distances using state discrimination", Phys. Rev. A **89**, 033861 (2014).